\documentclass[preprint, 3p, twocolumn, 12pt]{elsarticle}

\usepackage{amssymb}
\usepackage{array}
\usepackage{caption}
\usepackage{subcaption}
\usepackage{enumitem}
\usepackage{float}
\usepackage{xcolor}
\usepackage{multirow}

\usepackage[hyphens]{url}
\usepackage[hidelinks]{hyperref}
\hypersetup{breaklinks=true}

\usepackage{booktabs}

\journal{Energy and Built Environment}
\begin{document}
\begin{frontmatter}

\title{Toward Platform-based Building Design}




\author[inst1]{Yu-Wen Lin}
\author[inst2]{Tsz Ling Elaine Tang}
\author[inst3]{Stefano Schiavon}
\author[inst1]{Costas J. Spanos}

\affiliation[inst1]{organization={Department of Electrical Engineering and Computer Sciences, University of California, Berkeley},
            city={Berkeley},
            state={CA},
            country={USA}}

\affiliation[inst2]{organization={Siemens Technology},
            addressline={755 College Rd East}, 
            city={Princeton},
            postcode={08540}, 
            state={NJ},
            country={USA}}

\affiliation[inst3]{organization={Center for Built Environment, University of California, Berkeley},
            city={Berkeley},
            state={CA},
            country={USA}}
\begin{abstract}
The electronic design industry has undergone a significant transformation, transitioning from traditional hand-drawn designs to modern automated design processes. While Computer-Aided Design (CAD) tools emerged alongside the electronic industry, the current building design process has little to no automation. There is a need for a unified platform to address the complexity of building design and provide a more systematic approach.
Platform-based design (PBD), originally developed in the electronic industry, enables efficient design processes by promoting the reuse of hardware and software systems. It also facilitates design space exploration while optimizing performance.
This paper proposes a modular approach that divides the building into various disciplines and introduces a design flow using the PBD framework to streamline the design process.
We also present a case study that demonstrates the use of the PBD framework in the Heating, Ventilation, and Air Conditioning (HVAC) systems design. 
\end{abstract}


\end{frontmatter}


\section{Introduction}
\label{sec:intro}

Building accounts for more than 40\% of energy consumption in the United States \cite{USEIAEnergy2022}. 
A better building design can help in creating energy-efficient buildings that require less energy to heat, cool, and operate.
This can be achieved through the use of passive design strategies such as shading, insulation, and natural ventilation.
The orientation, shape, and shading of a building impact the amount of radiation that its surface is exposed to, potentially leading to an increase in cooling needs \cite{valladares2017review}.
The building envelope, such as roofs, walls, and doors, is another factor that plays an important role in regulating the internal temperature \cite{maniouglu2006economic}. 
For example, a well-insulated roof can help to reduce heat loss during the winter months, while walls with high thermal mass can help to absorb and store heat during the day and gradually release it at night, helping to maintain a comfortable internal temperature.
Additionally, properly sealed doors and windows can prevent drafts and heat loss, improving the overall energy efficiency of the building.
In recent decades, researchers have also explored the application of Phase Change Materials (PCMs) in building envelopes as a promising approach for thermal energy storage in buildings \cite{wang2009review}.
As a result, optimizing building envelope parameters can minimize heating load and improve energy efficiency.
Other passive techniques include ventilation, nocturnal convective cooling, radiant cooling, direct and indirect evaporative cooling \cite{givoni1991performance}.
These methods offer sustainable and energy-efficient ways to cool buildings without relying heavily on mechanical cooling systems.

Sustainable building materials can further reduce carbon emissions, making them a crucial aspect of modern building design. 
GHG labeling \cite{wu2014past} is increasingly important as it provides stakeholders with information on the amount of carbon emitted during material production and transportation, allowing designers to make informed decisions about the environmental impact of materials used at the early stages of design.
By selecting appropriate materials, designers can considerably reduce the environmental impact of a building.
In addition, green building certification programs, such as LEED and BREEAM, encourage the adoption of sustainable practices and help building owners and operators makes informed decisions to reduce energy consumption and carbon emissions.

However, the current state of the building design process is often fragmented, leading to sub-optimal designs.
The lack of automation in building design limits the exploration of innovative design options and hinders the efficiency of the overall process.
Automation has the potential to revolutionize building design by enabling the development of better design options. 
This paper explores some current advancements in automation for building design, focusing on three key areas: building structure and layout, Mechanical, Electrical, and Plumbing (MEP) systems, and building construction. 
It discusses the existing automation efforts in these disciplines and highlights their significance.
Additionally, we propose a module-based approach to building design, which enhances the organization and categorization of existing techniques.
By adopting this approach, it becomes easier to identify and utilize relevant automation methods for specific design tasks.
Furthermore, the paper introduces a three-layer PBD framework to facilitate a more comprehensive and detailed design flow within each discipline.
This framework allows for efficient exploration of design alternatives and optimization of performance.
A case study focusing on the design of HVAC systems is also presented to illustrate the proposed design flow.

\section{Background}
\label{sec:background}

Traditional design approaches are often limited by the experiences and knowledge of human designers.
Key design decisions are often already made before assessment due to budget, time constraints, and project requirements.
In addition, design evaluations primarily serve to meet building codes such as energy standards, ASHRAE 90.1 \cite{halverson2014ansi}, and thermal environmental conditions, ASHRAE 55 \cite{ashrae55}.  
Building design has the potential to be further optimized in order to create a more energy-efficient and environmentally-friendly solution.
Advances in computing technologies and machine learning algorithms have enabled the possibility of exploring a larger design space that may be overlooked by designers.
For example, Generative Adversarial Network (GAN) can be used to generate new building designs \cite{sun2022automatic, wu2022generative}, Genetic Algorithms (GAs) can be used to optimize building parameters and configurations with an objective of energy efficiency and thermal comfort \cite{tuhus2010genetic, li2017genetic}, and Natural Language Processing (NLP) algorithms can analyze building codes and regulations \cite{zhang2021deep, ding2022applications}. 
Existing BIM software, such as Autodesk Revit 2021 \cite{autodesk2021}, has the capacity to implement Generative Design (GD) for building geometry. 
It allows users to input design goals and constraints, and employs GD techniques to generate multiple design options to meet those objectives.  
It can be used to optimize building designs for structural performance, energy efficiency, and sustainability. However, limitations exist in the user interface, data availability, complex structures, and processing speed.

By utilizing advanced computational tools, designers can generate a set of high-quality solutions that meet design objectives and constraints, such as energy efficiency, functional performance, and cost-effectiveness.
The process can help accelerate the design processes, reduce costs, and improve the overall quality of building design. 
The following section reviewed prior studies of design optimization and automation in different disciplines of building design including building structure and layout, MEP systems, and construction. 

\subsection{Building Structure and Layout}
The building design process starts with an understanding of the project goals, scope, and budget. 
Then the design team conducts a preliminary analysis of the building site and explores different design options. 
Once the architect proposes a design concept for the building, which is then reviewed by the structural engineer to ensure that it meets requirements for structural feasibility and compliance with relevant building codes.
The communication between architects and engineers involves back-and-forth communications to arrive at a final design decision that meets both the aesthetic and functional requirements of the building. 
There is little to no real-time feedback when architects design the building. 
The back-and-forth communications between architects and engineers can be reduced through dynamic feedback on design tools.
With real-time feedback, architects can quickly evaluate the impact of their design choices on structural soundness, carbon footprint, energy efficiency, and cost-effectiveness.
Furthermore, if designers can get recommendations on potential design solutions with a better energy or carbon perspective, it not only saves design effort by improving the design pipeline but also creates a more sustainable design.
While the role of structural engineer can't be fully replaced, machine learning can assist in  performing structural analysis, optimizing building parameters and functional performance, and estimating cost. 

Building structural design consists of form, floor plan, fa\c{c}de, and structural components such as acoustic design and fire-resistant integration. 
The structural systems are responsible for energy and carbon emissions through the production and assembly of structural materials \cite{akbarnezhad2017estimation}.
As a result, it is crucial to take carbon emissions into account during construction design and material selections. 
Hammond et al. \cite{hammond2008embodied} created a database of 200 construction materials with their carbon and energy emission values. 
The database can facilitate estimating the carbon intensity of a building structure at the early stage of building design.
However, existing databases are sparse and are often constrained by building types, locations, and production companies.
To address the issue, Weber et al. \cite{weber2020generative} use a generative design approach to perform carbon analysis of steel framing systems based on building massing with 17\% error margin compared to actual buildings. 
Rather than providing an early-stage carbon assessment of a building, other research used optimization techniques to identify optimal building structural form.

The term "building structural form" refers to the overall shape, configuration, and style of a building. 
It involves the design of the physical appearance of a building.
Computer-aided Design and Engineering (CAD/E) software has the capability to automatically generate different options for building forms based on design constraints \cite{zboinska2015hybrid}.
Additionally, GA is also used to explore a wide range of design options and find optimal building forms based on energy performance \cite{yi2009optimizing, caldas2008generation, turrin2011design}.
Zhang and Blasetti applied GAN to study a style transfer of two buildings to inspire architects in various building forms \cite{zhang20203d}.

Once the structural form is established, the floor plan is developed. 
Designers sought automation during layout design for the purpose of faster design, design exploration, visual comforts, material effectiveness, and optimization \cite{weber2022automated}. 
The algorithms take building area or boundary as an input and output the program layout. 
The existing optimization methods include physically based \cite{de2016genetic}, evolution algorithm \cite{wong2009evoarch}, and Simulated Annealing (SA)  \cite{feng2016crowd}.  
However, challenges still exist in encoding social equity into a machine learn-able language.
As a result, floor plan generation methods at the current stage can't replace experienced human designers completely, but rather, serve as a guide to the designer.

Fa\c{c}ade design is another important element in the architectural design of a building. 
A fa\c{c}ade is the exterior wall or face of a building. 
It is pivotal not only in the aesthetic aspect of a building but also in energy efficiency, weather resistance, thermal and acoustic insulation, and lighting regulation.
Wang et al. \cite{wang2007facade} used a parametric optimization method to discover optimal fa\c{c}ade design of naturally ventilated residential buildings for better thermal comfort and energy savings. 
Torres and Sakamoto \cite{torres2007facade} determined optimal fa\c{c}ade design for daylight performance in a building using GA. 
Weber et al. \cite{weber2022solar} explored exoskeleton fa\c{c}ade design with parametric models based on different shading strategies for passive solar gain control and embodied carbon of materials.
GAN is also used in generating a visualization of fac\c{c}ade with existing fac\c{c}ade design image as training inputs \cite{newton2019generative}. 
Though the mentioned studies showed promising results in achieving a better design of fa\c{c}ade, most research didn't consider all objectives when forming an optimization problem. 
If all factors, such as aesthetic, energy, thermal, acoustic, and daylight, are considered, the formulated problem may become infeasible. 
The trade-offs between different objectives are crucial in the design decision.

Structural components are physical elements that make up the building's structure, such as beams, columns, walls, slabs, and foundations. 
At this stage, the process entails the selection of structural components that meet design constraints. 
Adeli and Park \cite{adeli1995optimization} applied a neural dynamics technique to design space trusses with the objective of minimizing the materials' weight. 
Bennage and Dhingra \cite{bennage1995single} used SA to optimize cross-sectional areas of trusses. 
Garrett and Fenves \cite{garrett1987knowledge} proposed a knowledge-based method to generate the detailed design of structural components. 
Caldas and Norford \cite{caldas2002design} studied GAs on optimal or near-optimal selections of construction materials.
Due to the availability of various construction materials, there could be several ideal solutions that meet the requirements for environmental performance. Nevertheless, the final design decision still requires the expertise of designers. 
Although the approach can offer useful information to designers, certain limitations exist, such as high computational expenses for a bigger design area, infeasible solutions, and the absence of a comprehensive global construction materials database.

\subsection{MEP Systems}
MEP systems are the next integral part of building design.
The mechanical systems include HVAC systems that regulate temperature, humidity, and air quality in the building.
The electrical systems include lighting, power distribution, and fire alarm systems. 
The plumbing systems include water supply, drainage, and waste management systems. 

Optimizing the design of HVAC systems is crucial for achieving energy efficiency in buildings, as these systems are responsible for using over 50\% of the electrical energy \cite{chua2013achieving}.
One challenge in the current design of HVAC systems is that the system is often oversized due to design safety factors and a lack of accurate building load calculations \cite{djunaedy2011oversizing}.
Oversizing the HVAC units has a major impact on energy inefficiency, cost, and thermal comfort. 
Trace 700 \cite{trace} is currently the most widely used HVAC sizing tool in the industry.
Once the system is sized, little to no automation exists in the industry during the HVAC design process.
Research studies have attempted to automate the design process of HVAC systems in component selection, topology, and duct design.
Zhang et. al. developed an automatic HVAC configuration of a two-zone model using GA \cite{zhang2005synthesis}.
The problem is formulated by optimizing three sub-systems: the decision of system type, the selection of components, and the choice of control strategies. 
However, one of the existing recommended HVAC configurations developed by ASHRAE \cite{ashrae2021handbook} still outperforms the resulting synthesis through optimization.
Another limitation is that the optimization problem may become infeasible for a larger system than the provided example. 
Asiedu et al. also use GA to optimize HVAC air duct system design with the objective of minimizing the life-cycle cost \cite{asiedu2000hvac}.
While achieving complete automation in HVAC design is not feasible at present due to the lack of standardization in HVAC design practices across different regions and non-uniformity in HVAC components and pricing, machines can assist in performing repetitive tasks such as duct routing and offering suggestions to designers.
Furthermore, the rule-based approach used in HVAC design guidelines presents a potential avenue for future design automation in the field.

The coordination between the three systems also poses a challenge during the MEP system design. 
Lu et al. \cite{lu2018mepcor} proposed a rule-based engine via Revit to facilitate the routing of the three systems. 
Designing and integrating MEP systems into a building is critical to ensure safety, thermal comfort, and functionality for its occupants. 
Furthermore, proper design and installation of MEP systems can also result in energy savings, lower operating costs, and carbon reduction.

\subsection{Building Construction}
BIM systems are often not fully utilized in the construction industry, even though the construction process is recorded in BIM, tremendous amounts of time and money can be saved.
One reason that this hasn't been done is that BIM requires specific hardware specs for full functionalities and the equipment is often challenging at the construction site.
Although the industry is working toward creating more agile apps for on-site applications in construction, the functionalities for data exchange and management are still limited.
Another reason is that manually inputting construction progress is tedious and time-consuming. 
However, with the advance in Artificial Intelligence (AI) and Internet of Things (IoT) systems, automatic data collection on-site during construction becomes possible. 
Technologies that have been applied to the automated data acquisition on the construction site include enhanced IT, geo-spatial positioning systems, 3D imaging, and augmented reality \cite{omar2016data}.
Asadi et al. \cite{asadi2020constructdata} proposed utilizing Unmanned Aerial/Ground Vehicles (UAV/UGV) to map the construction environment for accessing the current progress.
Zhang et al. \cite{zhang2022BIM-robot} explored automation between BIM and robotics in the construction process.
They proposed a framework for re-standardizing information flow during the construction process to enable automation for planning and executions.
Cheng et al. \cite{cheng2013real} used real-time positioning sensors to monitor dynamic resources such as personnel, equipment, and materials for safety during field operations.

Other than using AI or IoT technologies, in recent decades, the concept of modular buildings has been prevalent in the construction industry due to its higher efficiency and productivity.
Modular building is a concept that fabricates modules off-site as units and assembles the units on-site \cite{generalova2016modular}.
The benefits include reducing construction time, producing less waste, causing less disruption in the  surroundings, reducing labor requirements, and providing more flexibility for dynamic building layout adjustments during the progress \cite{bock2015future}.
Research has shown that with a modular design of a building, the construction process can become greatly streamlined in planning, off-site prefabrication, transportation, and on-site assembly with minimal human interventions \cite{balaguer2002futurehome}.
There are still limitations and obstacles for the current construction industry to become fully automated. 
Transitioning from a well-developed traditional construction practice to an AI-based modular approach requires additional risk, cost, and safety assessments.
In addition, design guidelines for building modules need to be standardized for enabling wide applications of the pre-fabricated modular structures.
Standardized guidelines allow architects and engineers to expand modular libraries to further create flexibility in the design process.
Lastly, coordination and management of robots on construction sites are still a challenging task \cite{gharbia2019robotic}. 
Though full automation in construction is still at an early stage in development, existing technologies show promising results in realizing automated construction.  

\section{Module-based Building Design}
\label{sec:module}

We propose a modular design flow that segregates various disciplines in building design into separate modules, aimed at facilitating the development of libraries and enabling design automation.


\begin{figure*}[htbp]
    \centering 
    \includegraphics[width=1.2\linewidth, angle=90]{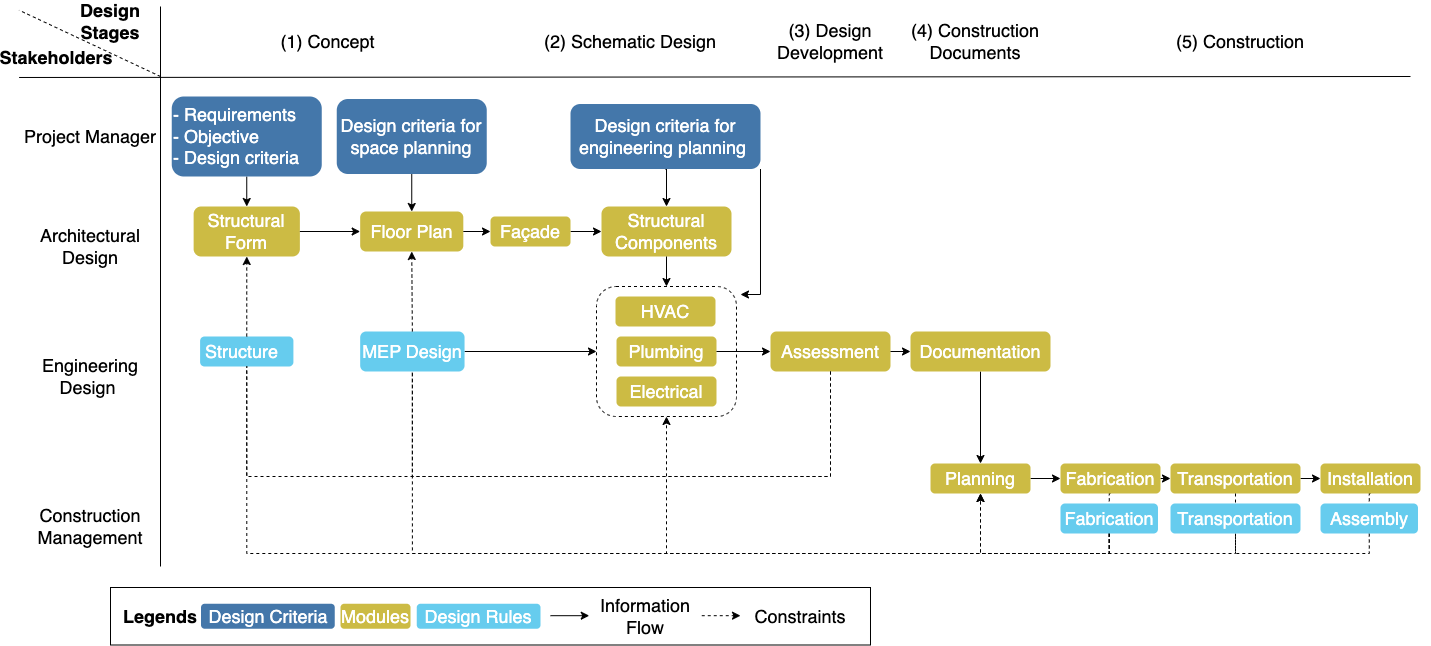}
    \caption{Proposed module-based approach for design automation in the building design process.}
    \label{fig:designprocess}
\end{figure*}

The building design process is generally split into five different design stages: concept, schematic design, design development, construction documents, and construction.
However, no clear objectives are defined at each stage. 
The ambiguity in defining different stages of building design arises from a lack of standardization, variation in project scope and complexity, and overlapping and iterative nature of the design process.
Due to the ambiguous definitions and varying inputs and outputs across different research groups, research in different disciplines of building design has become fragmented, making it difficult to integrate research into current industry practices. 
Identifying the gaps between research and the building industry is therefore essential to bring them closer together and bridge the divide. 
Our aim is to minimize these gaps and facilitate the transfer of knowledge and research findings to the industry.

Fig \ref{fig:designprocess} describes the proposed module-based approach of the design process that integrates design automation techniques of different disciplines into a design flow.
Different disciplines of the building design process are split into different modules: structural form, floor plan, fa\c{c}ade, structural components, HVAC, electrical, plumbing, assessment, documentation, planning, fabrication, transportation, and installation.
Each module contains design automation methods with specific input-output specifications as shown in Table \ref{table_archmodule} and \ref{table_ecmodule}.
The objective is not to create a comprehensive review of design automation studies, but to offer instances of commonly employed techniques in research communities.
Modules are connected to one another to demonstrate the design sequence and they are bounded by design rules.
The design rules are generally established through physical constraints, building codes, safety standards, environmental regulations, and budget constraints. 

The building design process commences by creating the structural form of the building based on client requirements, building objectives, and design criteria for the site, while adhering to structural constraints and energy building codes.
Then, the floor plan is developed by integrating inputs from the structural form and space planning design criteria.
However, it is subject to MEP design rules to ensure sufficient space for MEP equipment. 
Following the floor plan is designed, fa\c{c}ade and structural components are determined to finalize the architectural design of the building.
Subsequently, HVAC, electrical, and plumbing systems are designed in a coordinated manner to meet the occupants' needs and spatial requirements.  
Once the MEP system is in place, an assessment module performs a final check of the building code before creating the construction documents. 
The construction planning module then facilitates planning for fabrication, transportation, and installation during the construction stage.
Lastly, those modules in the construction stage implement the construction plan to complete the building design. 

Although the figure shows that the modules are linked, there is still a discrepancy between the outputs of one module and the inputs of the connected module.
Currently, the linked connection requires manual intervention to ensure a proper information flow between modules, and the backbone of the information flow is BIM. 
Several researchers have been working on extending the data structure, such as Industry Foundation Classes (IFC) \cite{IFC_iso201816739}, to fit different applications' needs \cite{froese2003future}.
However, the current data structure still hasn't addressed design automation applications completely.
Additionally, the proposed design flow is not a single iteration process from left to right.
Revisions or adjustments from previous modules may be necessary to ensure compatibility and proper functioning of the overall system. 
However, over time, design rules can be expanded by incorporating feedback from professional experts and by collecting more data and information to better address the needs and constraints of the project.
They can facilitate reducing the number of iterations needed to create an optimal building design.
Lastly, the proposed approach is not intended to replace architects and engineers completely, but rather, to offer design suggestions and enhance communication between designers and engineers during the design process.

The design process is designed to be flexible to accommodate regional, climatic, and firm-specific variations in the building design process. Furthermore, the design process encourages researchers to reconsider the necessity of standardizing information flow between different disciplines, allowing designers to easily compare and contrast the results of various methods.

\begin{table*}[htbp]
\caption{Architectural Design Modules}
\label{table_archmodule}
\centering
\scriptsize
\begin{tabular}{p{0.13\linewidth}  p{0.2\linewidth}  p{0.14\linewidth}  p{0.17\linewidth}  p{0.23\linewidth} }
    \toprule
    \multicolumn{2}{l}{\textbf{Module}} & \multicolumn{2}{l}{\textbf{Behavioral Model}} & \textbf{Extra-Functional Model} \\
    \textit{object} & \textit{Method} & \textit{Inputs} & \textit{Outputs} & \textit{Objective} \\
    \midrule \\
    \multirow{3}{*}{Structural Form} & CAD/E \cite{zboinska2015hybrid} & Design objectives and constraints & Building forms & N/A\\
    \rule{0pt}{4ex}
    & GD \cite{weber2020generative} & Building outline, loading, span & Structure model of steel framing systems and structural material quantities & Embodied Carbon \\
    \rule{0pt}{4ex}
    & GA \cite{yi2009optimizing} & Building size & Building forms & Energy performance \\
    \rule{0pt}{2ex} \\
    \hline \\
    \multirow{4}{*}{Floor Plan} & Physically based \cite{de2016genetic} & Spaces and shape & Space topology & Adjacency, separation, orientation, alignment, area, proportion \\
    \rule{0pt}{4ex}    
    & Evolution Algorithm \cite{wong2009evoarch} & Floor area, room ratios, adjacency preference matrix & Space topology & Budget, adjacency preferences and space function \\
    \rule{0pt}{4ex}    
    & Agent-based modeling, SA \cite{feng2016crowd} & Simulated occupants' behavior & Space layout & Mobility, accessibility, and coziness \\
    \rule{0pt}{4ex}    
    & Agent-based modeling, Linear/Quadratic Programming \cite{bisht2022transforming} & Building spaces & Optimal assignment of building spaces & Occupants' flow, satisfaction, and energy efficiency \\
    \rule{0pt}{2ex}   \\
    \hline \\
    \multirow{2}{*}{Fa\c{c}ade} & Parametric Model \cite{wang2007facade} & Building geometry & U-values of fa\c{c}ades & Thermal comfort and energy savings \\
    \rule{0pt}{4ex} 
    & GA \cite{torres2007facade} & Geometric parameters & Parameters' value & Daylight performance \\
    \rule{0pt}{4ex} 
    & GAN \cite{newton2019generative} & existing fac\c{c}ade design images & fac\c{c}ade design & N/A \\ 
    \rule{0pt}{2ex}   \\
    \hline \\
    \multirow{2}{0.13\linewidth}{Structural Components} & Knowledge-based Method \cite{garrett1987knowledge} & Building structure & Detailed design of structural components & N/A\\
    \rule{0pt}{4ex}   
    & Neural Network \cite{adeli1995optimization} & Loading condition & Cross-sectional areas of trusses & Material weight \\
    \rule{0pt}{4ex}    
    & SA \cite{bennage1995single} & Loading condition & Cross-sectional areas of trusses & Material Weight   \\
    \rule{0pt}{4ex}
     & GA \cite{caldas2002design} & Building structure & Construction materials properties & Thermal and lighting performance \\
    \rule{0pt}{2ex} \\
    \bottomrule
\end{tabular}
\end{table*}

\begin{table*}[htbp]
\caption{Engineering and Construction Design Modules}
\label{table_ecmodule}
\centering
\scriptsize
\begin{tabular}{p{0.13\linewidth}  p{0.2\linewidth}  p{0.14\linewidth}  p{0.17\linewidth}  p{0.23\linewidth} }
    \toprule
    \multicolumn{2}{l}{\textbf{Module}} & \multicolumn{2}{l}{\textbf{Behavioral Model}} & \textbf{Extra-Functional Model} \\
    \textit{object} & \textit{Method} & \textit{Inputs} & \textit{Outputs} & \textit{Objective} \\
    \midrule \\
    \multirow{2}{*}{HVAC} & Physics-based (BPS) \cite{barnaby2001hvac} &  Building structure and layout & Duct and radiant systems placement & Cost, GHG emissions, life-cycle \\
    \rule{0pt}{4ex}
    & GA \cite{zhang2005synthesis} & Building structure and layout & HVAC component selections and topology & Energy efficiency  \\
    \rule{0pt}{2ex} \\
    \hline \\
    Assessment & Rule-based \cite{ansah2019review} & BIM & Report documents & Green building assessment\\
    \rule{0pt}{2ex} \\
    \hline \\
    \multirow{3}{*}{Planning} & Computer Integral Manufacturing \cite{penin1998robotized} & BIM  & Off-site prefabrication plan & Time, transportation, market, legal, economic \\
    \rule{0pt}{4ex}
    & Mixed-Integer Programming \cite{almashaqbeh2022multiobjective} & BIM & Transportation plan (Truck assignment and delivery day) & Cost \\
    \rule{0pt}{4ex}
    & Imaged-based 3D modeling \cite{ding2020bim} & BIM, robot property, task & Assembly plan & Time \\
    \rule{0pt}{2ex} \\
    \hline \\
    \multirow{4}{*}{Data Collection} & UAV/UGV \cite{asadi2020constructdata} & Construction site & Construction environment map & Surveying, monitoring, and inspection \\
    \rule{0pt}{4ex}
    & RF/RFID \cite{navon2003monitoring} & Construction site & Labor locations & Project performance \\
    \rule{0pt}{4ex}
    & Positioning sensors \cite{cheng2013real} & Construction site & Dynamic resources quantities & Real-time resource availability \\
    \rule{0pt}{4ex}
    & 3D laser scanning \cite{akinci2006formalism} & Construction site & 3D point clouds & Quality Control \\
    \rule{0pt}{2ex} \\
    \bottomrule
\end{tabular}
\end{table*}

\section{Platform-based Building Design}
\label{sec:pbd}

In the previous section, we provided an overview of the building design flow at a higher level.
In this section, we delve into a more detailed view on how each module can be implemented using Platform-based Design (PBD).
PBD \cite{ferrari1999system} refers to an approach in the design paradigm where a common base or platform is used as a foundation for creating multiple variations or derivatives of a design.
It emphasizes the reuse of components across different designs, allowing for faster and more efficient product development.
While Electronic Design Automation (EDA) is a prime example, PBD has also been successfully applied in other industries, including automotive \cite{simpson2006platform}, aerospace \cite{becz2010design}, and healthcare \cite{yang2014health}. 
In PBD, a \textit{platform} is defined as a collection of components and their accompanying rules for combining them, which can be utilized to create a design at a specific level of abstraction.
Jia et al. \cite{jia2018design} proposed a PBD approach specifically tailored for smart building design. 
The approach focuses on promoting the reuse of hardware and software components on shared infrastructures.
They present a case study of retrofitting the HVAC system in a smart building by deploying sensors and actuators to enhance energy efficiency and occupants' comfort. 
The PBD approach enables the reuse of sensors and actuators across different smart building designs, leading to more cost-effective and sustainable solutions.
We follow a similar approach, but instead of applying it to building retrofit applications, we apply the technique to building design, specifically for the HVAC systems.

The approach follows closely to the design flow developed from \cite{jia2018design}, with the adaptations to suit the building design process.
The concept is to begin with high-level design specifications and gradually refine the model in subsequent steps, leading to the implementation of the design at different levels of abstraction.
This design flow does not strictly follow a top-down or bottom-up approach, but instead adopts a meet-in-the-middle approach that combines elements of both processes.
The bottom-up approach begins with detailed components or subsystems and builds up to a higher-level design.

\begin{figure*}[htbp]
    \centering
    \includegraphics[width=\linewidth]{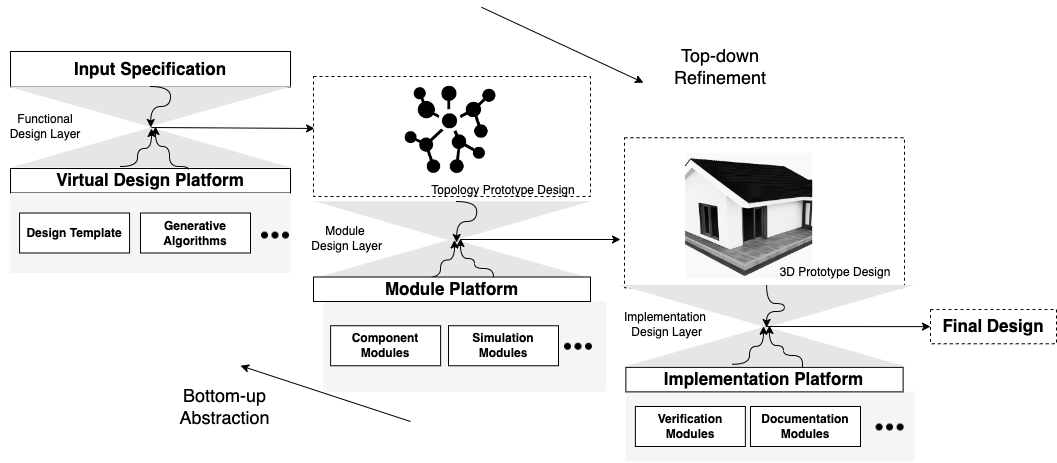}
    \caption{An overview of the proposed design flow adapted from \cite{jia2018design}.}
    \label{fig:pbd_overview}
\end{figure*}

The design flow is a multi-layered process that involves the functional design layer, the module design layer, and the implementation design layer, as depicted in Figure \ref{fig:pbd_overview}.
Each layer is accompanied by corresponding libraries that aid in the design process.
The libraries that correspond to each layer are a virtual design platform, a module platform, and an implementation platform, respectively.

The virtual design platform is composed of several components, including design templates and generative algorithms.
Design templates are created based on building codes or previous building projects, and provide a suggested layout for a specific design specification.
On the other hand, generative algorithms utilize AI to generate alternative design suggestions based on the input data and specifications.
The libraries enable designers to explore various design options and select the most suitable one based on the specifications, thus enabling them to advance to the next level of abstraction.
The functional design layer yields the topology prototype design, which depicts the interconnections between different components.
Subsequently, the module design layer comprises component and simulation modules.
This layer facilitates designers to experiment with different components and conduct simulation studies to evaluate hypothetical scenarios of the resulting design.
The outcome of this layer is a 3D Prototype design, which is a more detailed representation design. 
In the implementation platform, the design undergoes verification and documentation processes using verification and documentation modules.
These modules aid in verifying the design's functionality and producing documentation to describe the design's specifications.

\section{Case Study}

We present a case study to demonstrate the practical application of the PBD framework in the design of HVAC systems for a building.
Assuming we are given a BIM that contains building form and floor plan information, the design specification consists of the desired design cost, comfort requirements, and maximum energy consumption value.

In the functional design layer, the HVAC configuration templates developed by ASHRAE \cite{ashrae2021handbook} are utilized to match the design specifications.
These templates are selected based on the building type and typical heat load of the zones and the building.
Although researchers have attempted to develop new HVAC topologies using GA, Zhang et al. \cite{zhang2005synthesis} found that the existing configurations still outperform the new generative alternatives.
However, if future generative studies succeed in outperforming the existing configurations, they can be stored in the virtual design platform.
Once the appropriate configurations are selected, the component modules stored in the module platform are used to select the best-suited components for the design.
Additionally, 2D topology is converted to 3D by applying duct layout modules that fit the air ducting system to the building floor plan.
The resulting design is then sent to the implementation layer for implementation.
The implementation layer consists of a final verification stage, where the design is checked for compliance with building codes and feasibility for implementation.
Once validated, the documentation module is used to generate the final construction documents required for bidding purposes.

\begin{figure*}[htbp]
    \centering    
    \includegraphics[width=1.3\linewidth, angle=90]{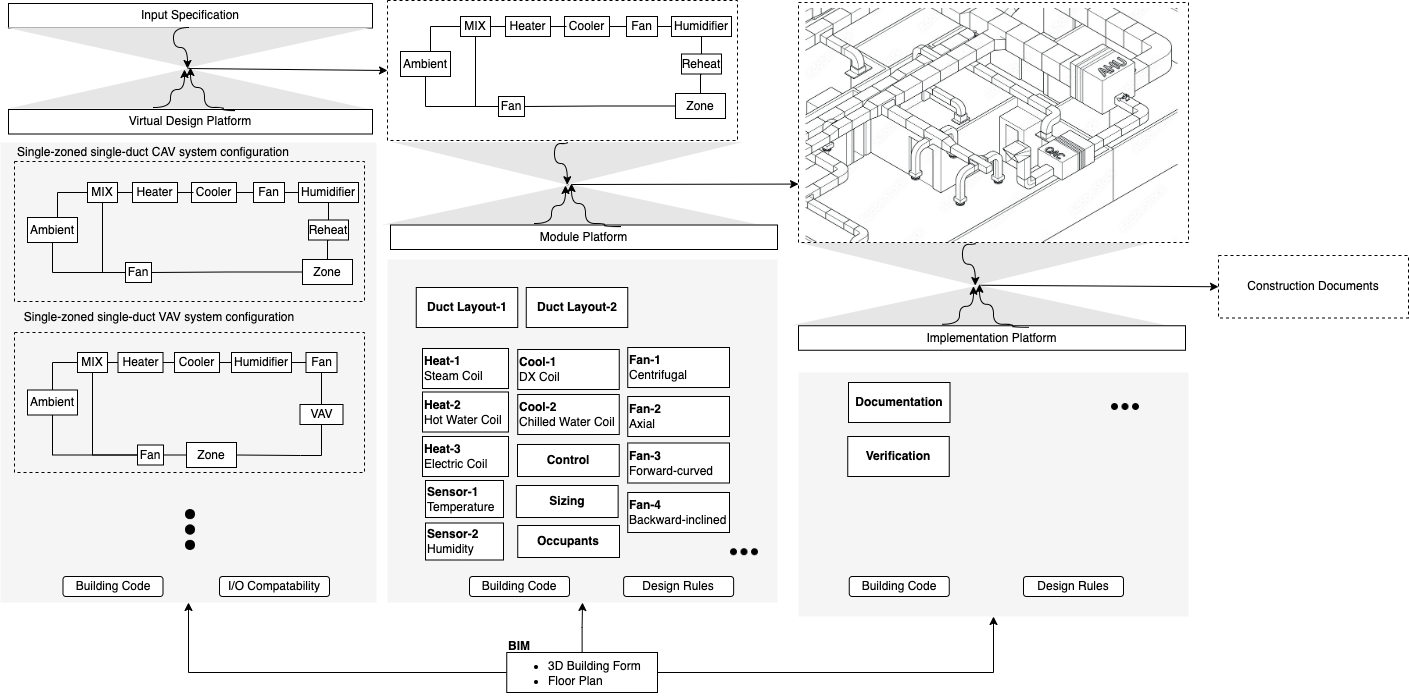}
    \caption{A detailed overview of the PBD design flow of an HVAC system.}
    \label{fig:hvac-pbd}
\end{figure*}
\section{Conclusion and Future Work}

In this paper, we examined the existing efforts in design automation in the filed of building design.
We proposed a high-level modular design flow, which separates different disciplines involved in building design into distinct modules.
The modular approach allows for better organization and management of the design process.
Then, we developed a more detailed design flow based on PBD that described the design flow of a specific discipline, providing a systematic approach to design.
To illustrate the practical application of the proposed framework, we provided a conceptual case study focused on an the HVAC system. 
The proposed methodology can also be applied to other discipline in building design such as structural form and floor plan generation.

However, the implementation of the PBD design flow does come with certain challenges that need to be addressed.
One of the primary challenges is the establishment of data and library standards for building components.
It is crucial to ensure that the input-output relationships of components within the same modules are consistent and compatible, enabling seamless interchangeability and testing across different design environments.
Moreover, to maintain the effectiveness of the design libraries, it is necessary to  populate them with the latest research findings and advancements in the field. This ensures that the libraries remain up-to-date and enable easy comparison and contrast with traditional design approaches.

Another challenge arises during the mapping process, specifically when mapping from one level of abstraction to the next.
Given the complexity and multi-objective nature of building design, the optimization problem may become infeasible.
Trade-offs must be carefully evaluated to determine the most viable design solutions that balance competing objectives effectively.
This evaluation process helps navigate the complexities of building design and ensures that the resulting designs are practical and well-suited to the intended purposes.

Addressing these challenges in future research is essential for adapting the building design flow to the PDB design framework.
This adaptation holds the potential to offer substantial benefits in terms of efficiency, collaboration, and innovation.
The framework facilitates a more streamlined process, which leads to the creation of more sustainable and optimized building designs.


 \bibliographystyle{elsarticle-num} 
 \bibliography{refs}
 
\end{document}